\documentclass[12pt]{iopart}
\usepackage{graphicx}
\begin{document}
\jl{03}

\title[Author guidelines for IOP Journals]
{A common behavior of thermoelectric layered cobaltites:
\\ incommensurate spin density wave states in 
[Ca$_2$Co$_{4/3}$Cu$_{2/3}$O$_4$]$_{0.62}$[CoO$_2$] 
and 
[Ca$_2$CoO$_3$]$_{0.62}$[CoO$_2$]
}

\author{J Sugiyama\dag\footnote[3]{To 
whom correspondence should be addressed.}, 
J H Brewer\ddag\, 
E J Ansaldo$\|$, H Itahara\dag, K Dohmae\dag, 
C Xia\dag, Y Seno\dag, B Hitti$\|$\  and T Tani\dag}

\address{\dag\ Toyota Central Research and Development Labs. Inc., 
 Nagakute, Aichi 480-1192, Japan}

\address{\ddag\ TRIUMF, CIAR and 
Department of Physics and Astronomy, University of British Columbia, 
Vancouver, BC, V6T 1Z1 Canada}

\address{$\|$\ TRIUMF, 4004 Wesbrook Mall, Vancouver, BC, V6T 2A3 Canada}

\begin{abstract}
Magnetism of a misfit layered cobaltite 
[Ca$_2$Co$_{4/3}$Cu$_{2/3}$O$_4$]$_x^{\rm RS}$[CoO$_2$] 
($x \sim$ 0.62, RS denotes a rocksalt-type block) was investigated 
by a positive muon spin rotation and relaxation ($\mu^+$SR) experiment. 
A transition to an incommensurate ({\sf IC}) spin density wave 
({\sf SDW}) state was found below 180~K (= $T_{\rm C}^{\rm on}$); 
and a clear oscillation due to a static internal magnetic field 
was observed below 140~K (= $T_{\rm C}$). 
Furthermore, an anisotropic behavior of 
the zero-field $\mu^+$SR experiment 
indicated that the {\sf IC-SDW} propagates in the $a$-$b$ plane, 
with oscillating moments directed along the $c$ axis. 
These results were quite similar to those for the related compound 
[Ca$_2$CoO$_3$]$_{0.62}^{\rm RS}$[CoO$_2$], {\sl i.e.}, Ca$_3$Co$_4$O$_9$. 
Since the {\sf IC-SDW} field in 
[Ca$_2$Co$_{4/3}$Cu$_{2/3}$O$_4$]$_{0.62}^{\rm RS}$[CoO$_2$] 
was approximately same to those in pure and doped
[Ca$_2$CoO$_3$]$_{0.62}^{\rm RS}$[CoO$_2$],
it was concluded that the {\sf IC-SDW} exist in the [CoO$_2$] planes. 
\end{abstract}



\maketitle

\section{Introduction}
Layered cobaltites are investigated eagerly 
because of their structural and compositional variety 
and also their good thermoelectric properties.  
At present, the following three groups of cobaltites are known 
to be good thermoelectrics, because they display 
metallic conductivities as well as high thermoelectric powers $S$, 
for reasons which are currently not fully understood.
A sodium cobaltite, Na$_x$CoO$_2$, 
was the first compound reported as a good thermoelectric material
\cite{NCO_1,NCO_2,NCO_3}.
Then, the finding of 
[Ca$_2$CoO$_3$]$_{0.62}^{\rm RS}$[CoO$_2$]
\cite{CCO_1,CCO_2,CCO_3}, 
and then
[Sr$_2$Bi$_{2-y}$Pb$_y$O$_4$]$_x^{\rm RS}$[CoO$_2$]
\cite{4LBiSrCO_1,4LBiSrCO_2,4LBiSrCO_3},
followed \cite{4LCCCO_1}, 
where RS denotes a rocksalt-type block.
All share a common structural component, 
the CoO$_2$ planes, in which a two-dimensional-triangular lattice 
of Co ions is formed by a network of edge-sharing CoO$_6$ octahedra. 
Charge carrier transport in these cobaltites is thought to be 
restricted mainly to the CoO$_2$ planes, 
as in the case of the CuO$_2$ planes for the high-$T_c$ cuprates. 

Recent positive muon spin rotation and relaxation ($\mu^+$SR) 
experiments on 
[Ca$_2$CoO$_3$]$_{0.62}^{\rm RS}$[CoO$_2$]
\cite{jun1_3L,jun2_3L,jun3_3L,jun4_3L} 
indicated the existence of an incommensurate ({\sf IC}) 
spin density wave ({\sf SDW}) state below 100~K, 
which was not detected previously by other magnetic 
measurements \cite{CCO_2,CCO_3,jun1_3L}. 
The latter two experiments suggested that 
a long-range {\sf IC-SDW} order completed below $\sim$ 30~K, 
while a short-range order appeared below 100~K\cite{jun3_3L,jun4_3L}. 
Since the $\rho(T)$ curve exhibits a broad minimum 
around 80~K \cite{CCO_2,CCO_3,jun1_3L}, 
the behavior of conduction electrons is found to be 
strongly affected even by their short-range magnetic order.

A new thermoelectric layered cobaltite 
[Ca$_2$Co$_{4/3}$Cu$_{2/3}$O$_4$]$_{0.62}^{\rm RS}$[CoO$_2$] 
was found recently \cite{4LCCCO_1};
the crystal structure 
consists of alternating layers of the quadruple rocksalt-type 
[Ca$_2$Co$_{4/3}$Cu$_{2/3}$O$_4$] subsystem 
and the single CdI$_2$-type [CoO$_2$] subsystem 
stacked along the $c$ axis. 
There is a misfit between these subsystems along the $b$-axis, 
similar to the case of [Ca$_2$CoO$_3$]$_{0.62}^{\rm RS}$[CoO$_2$] 
\cite{CCO_2,CCO_3}.
Polycrystalline 
[Ca$_2$Co$_{4/3}$Cu$_{2/3}$O$_4$]$_{0.62}^{\rm RS}$[CoO$_2$] samples
have values of thermopower $S$ = 150~$\mu$VK$^{-1}$ and 
resistivity $\rho$ =15~m$\Omega$cm at 300~K \cite{4LCCCO_1}. 
As a result, their thermoelectric power factor (= $S^2\rho^{-1}$) is 
$\sim$ 20\% larger than that of polycrystalline 
[Ca$_2$CoO$_3$]$_{0.62}^{\rm RS}$[CoO$_2$] samples. 
Since the $\rho(T)$ curve exhibited a broad minimum and 
the $S(T)$ curve a broad maximum around 130~K, 
there seems to exist a transition from 
a high-temperature metallic to a low-temperature insulator state 
around 130~K. 
On the other hand, susceptibility ($\chi$) measurements indicate 
no anomalies around 130~K, 
although $\chi$ showed a small change at $\sim$ 80~K.

Therefore, $\mu^+$SR experiments on 
[Ca$_2$Co$_{4/3}$Cu$_{2/3}$O$_4$]$_{0.62}^{\rm RS}$[CoO$_2$] 
are also expected to provide crucial information on the correlation 
between magnetism and transport properties 
in the layered cobaltites. 
Furthermore, such experiment is significant to 
clarify the universal behavior of magnetism 
in thermoelectric layered cobaltites. 
Here, we report both weak ($\sim$ 100~Oe) transverse-field (wTF-) $\mu^+$SR 
and zero field (ZF-) $\mu^+$SR measurements 
for a $c$ axis aligned 
[Ca$_2$Co$_{4/3}$Cu$_{2/3}$O$_4$]$_{0.62}^{\rm RS}$[CoO$_2$] 
sample at temperatures below 300~K.  
The former method is sensitive to local magnetic order 
{\it via\/} the shift of the $\mu^+$ spin precession frequency 
and the enhanced $\mu^+$ spin relaxation, 
while ZF-$\mu^+$SR is sensitive to weak local magnetic [dis]order 
in samples exhibiting quasi-static paramagnetic moments.

\section{Experiment}

A $c$-axis aligned polycrystalline 
[Ca$_2$Co$_{4/3}$Cu$_{2/3}$O$_4$]$_{0.62}^{\rm RS}$[CoO$_2$] plate 
($\sim 15 \times 10 \times 2$~mm$^3$) 
was synthesized by a reactive templated grain growth technique \cite{rtgg_1}. 
In addition, $c$-axis aligned cobaltites with the triple rocksalt-type subsystem, 
{\sl i.e.}, 
[Ca$_2$CoO$_3$]$_{0.62}^{\rm RS}$[CoO$_2$] 
and 
[Ca$_{1.8}M_{0.2}$CoO$_3$]$_x^{\rm RS}$[CoO$_2$] 
($M$ = Sr, Y and Bi)
plates 
were prepared for comparison.
Then, the $c$-aligned plates were annealed at 450~$^o$C 
for 12 hours in an oxygen flow.
Powder X-ray diffraction ({\sl XRD\/}) studies indicated that 
all the samples
were single phase with a monoclinic structure and almost 
100\% aligned along the $c$ axis.  
The preparation and characterization of these samples 
were described in detail elsewhere \cite{rtgg_2}.  

Magnetic susceptibility ($\chi$) was measured using 
a superconducting quantum interference device ({\sl SQUID\/}) 
magnetometer ({\sf mpms}, {\it Quantum Design\/}) 
in a magnetic field of less than 55~kOe.  
The $\mu^+$SR experiments were performed on the 
{\bf M15} or {\bf M20} surface muon beam lines at TRIUMF.  
The experimental setup and techniques are described elsewhere \cite{ICSDW_1}.  

\section{Results}
\subsection{
[Ca$_2$Co$_{4/3}$Cu$_{2/3}$O$_4$]$_{0.62}^{\rm RS}$[CoO$_2$] 
}

\begin{figure}
\begin{center}
\includegraphics[width=8cm]{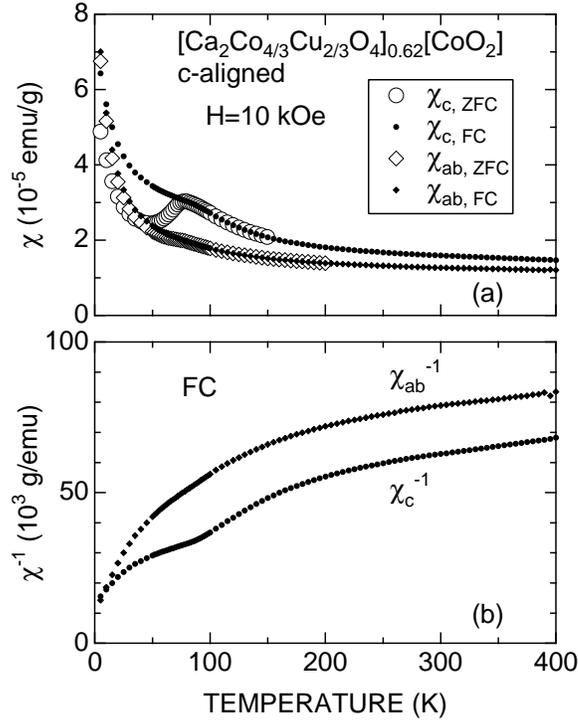}
\caption{\label{fig:chi}Temperature dependences of 
(a) susceptibility $\chi$ and (b) $\chi^{-1}$ for 
$c$-aligned [Ca$_2$Co$_{4/3}$Cu$_{2/3}$O$_4$]$_{0.62}^{\rm RS}$[CoO$_2$].
}
\end{center}
\end{figure}

Figures~\ref{fig:chi}(a) and \ref{fig:chi}(b) show 
the temperature dependences of $\chi$ and $\chi^{-1}$ for the $c$-aligned 
[Ca$_2$Co$_{4/3}$Cu$_{2/3}$O$_4$]$_{0.62}^{\rm RS}$[CoO$_2$] sample. 
In order to determine anisotropy, the magnetic field $H$ was applied 
parallel or perpendicular to the {\sl ab} plane. 
Hereby, we abbreviate $\chi$ obtained with $H \bot ab$ as $\chi _c$ 
and $H // ab$ as $\chi _{ab}$, respectively. 
The $\chi _c(T)$ curve in a zero-field cooling ({\sf ZFC}) mode exhibits 
a cusp at $\sim$ 85~K; 
also, a clear thermal hysteresis is seen 
between the data obtained in a {\sf ZFC} mode and 
a field cooling ({\sf FC}) mode. 
On the other hand, as $T$ decreases, $\chi _{ab}^{-1}$ decreases 
monotonically with increasing slope (d$\chi _{ab}^{-1}$/d$T$), 
although there is a small anomaly at $\sim$ 85~K, 
probably due to a misalignment between the sample axis and $H$. 
Nevertheless, the magnetization ($M) - H$ curve did not show 
a clear loop even at 5~K. 
These results suggest that 
[Ca$_2$Co$_{4/3}$Cu$_{2/3}$O$_4$]$_{0.62}^{\rm RS}$[CoO$_2$] 
undergoes a magnetic transition to 
either a ferrimagnetic or a spin glass state at 80~K 
with the easiest magnetization direction parallel to the $c$-axis. 
It is worth noting that there are no marked anomalies in Figure~\ref{fig:chi} 
except for the cusp in the $\chi _c(T)$ at $\sim$ 85~K. 

\begin{figure}
\begin{center}
\includegraphics[width=8cm]{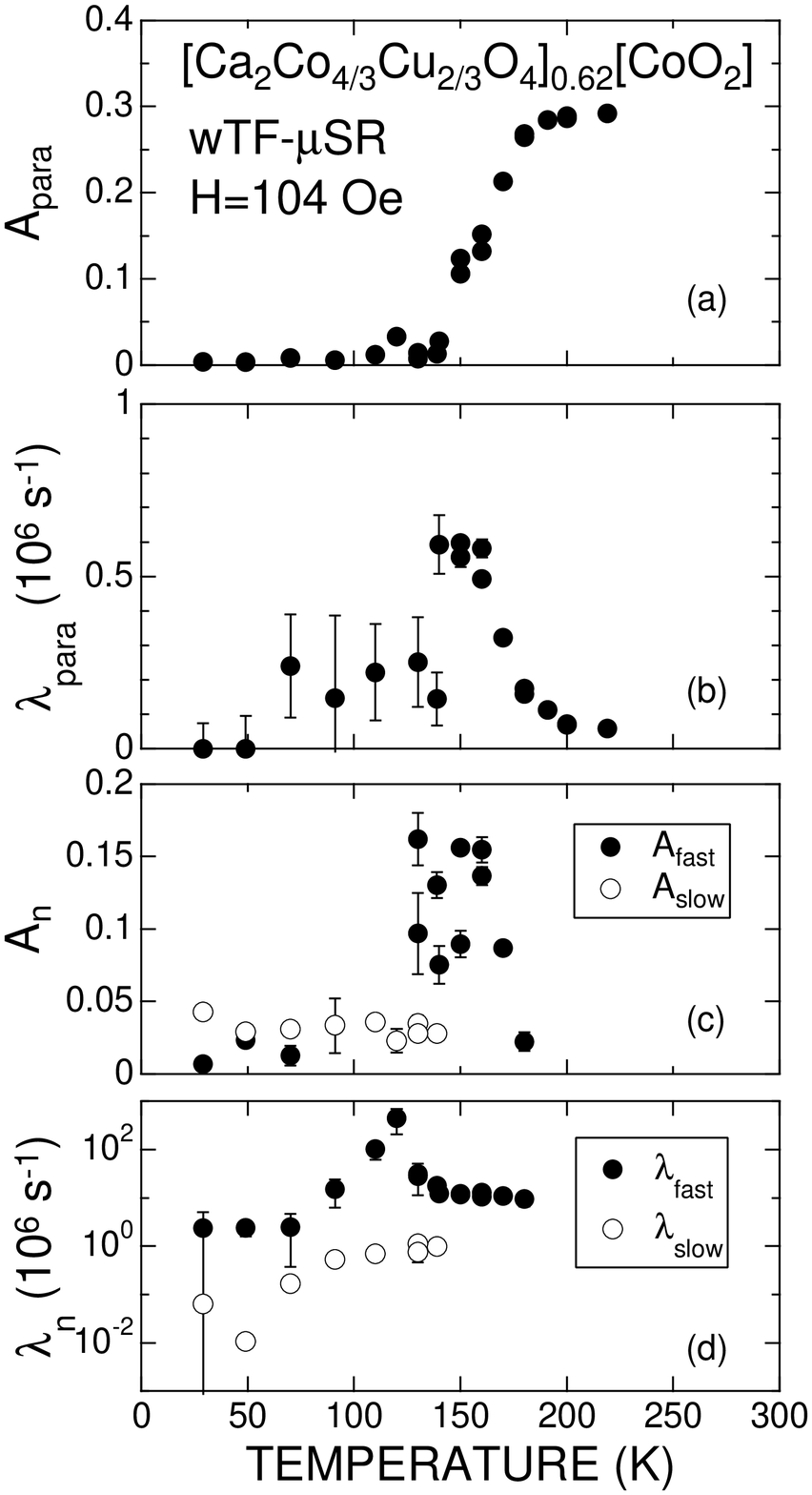}
\caption{\label{fig:wTF-muSR}Temperature dependences of 
(a) $A_{\sf para}$, (b) $\lambda_{\sf para}$, (c) $A_{n}$ and 
(d) $\lambda_{n}$ ($n$ = {\sf fast} and {\sf slow}) in 
$c$-aligned [Ca$_2$Co$_{4/3}$Cu$_{2/3}$O$_4$]$_{0.62}^{\rm RS}$[CoO$_2$]. 
The data were obtained by fitting the wTF-$\mu^+$SR spectra 
using equation (\ref{eq:TFfit}).
}
\end{center}
\end{figure}

The wTF-$\mu^+$SR spectra in a magnetic field of $H \sim100$~Oe 
in the $c$-aligned 
[Ca$_2$Co$_{4/3}$Cu$_{2/3}$O$_4$]$_{0.62}^{\rm RS}$[CoO$_2$] sample
exhibit a clear reduction of the $\mu^+$ precession amplitude below 200~K.  
The wTF-$\mu^+$SR spectrum below 200~K was well fitted in the time domain with a combination of 
 a slowly relaxing precessing signal and 
 two non-oscillatory signals, one fast and the other slow relaxing:
\begin{eqnarray}
A_0 \, P(t) &=& A_{\sf para} \, \exp(- \lambda_{\sf para} t) \, \cos (\omega_\mu t + \phi)
\cr
&+& A_{\sf fast} \, \exp(-\lambda_{\sf fast} t) 
\cr
&+& A_{\sf slow} \, \exp(-\lambda_{\sf slow} t), 
\label{eq:TFfit}
\end{eqnarray}
where $A_0$ is the initial asymmetry, 
$P(t)$ is the muon spin polarization function, 
$\omega_\mu$ is the muon Larmor frequency, 
$\phi$ is the initial phase of the precession and 
$A_n$ and $\lambda_n$ ($n$ = {\sf para}, {\sf fast} and {\sf slow}) 
are the asymmetries and exponential relaxation rates of the three signals.  
The latter two signals ($n$ = {\sf fast} and {\sf slow}) 
have finite amplitudes 
below $\sim 180$~K.

Figures~\ref{fig:wTF-muSR}(a) - \ref{fig:wTF-muSR}(d) 
show the temperature dependences of 
$A_{\sf para}$, 
$\lambda_{\sf para}$, 
$A_{n}$ ($n$ = {\sf fast} and {\sf slow}) and 
$\lambda_{n}$ 
in the $c$-aligned 
[Ca$_2$Co$_{4/3}$Cu$_{2/3}$O$_4$]$_{0.62}^{\rm RS}$[CoO$_2$] sample.
The large decrease in $A_{\sf para}$ below 180~K 
(and the accompanying increase in $\lambda_{\sf para}$) 
indicate the existence of a magnetic transition 
with an onset temperature $T_c^{\rm on} \sim 180$~K, 
a transition width $\Delta T \sim 40$~K and 
an endpoint $T_c^{\rm end} \sim 140$~K, respectively.
Since $A_{\sf para}$ is roughly proportional to 
the volume of a paramagnetic phase, 
this result ($A_{\sf para} \sim$ 0 below $T_c^{\rm end}$) suggests that 
almost all the sample changes into a magnetically ordered state below 140~K. 

\begin{figure}
\begin{center}
\includegraphics[width=8cm]{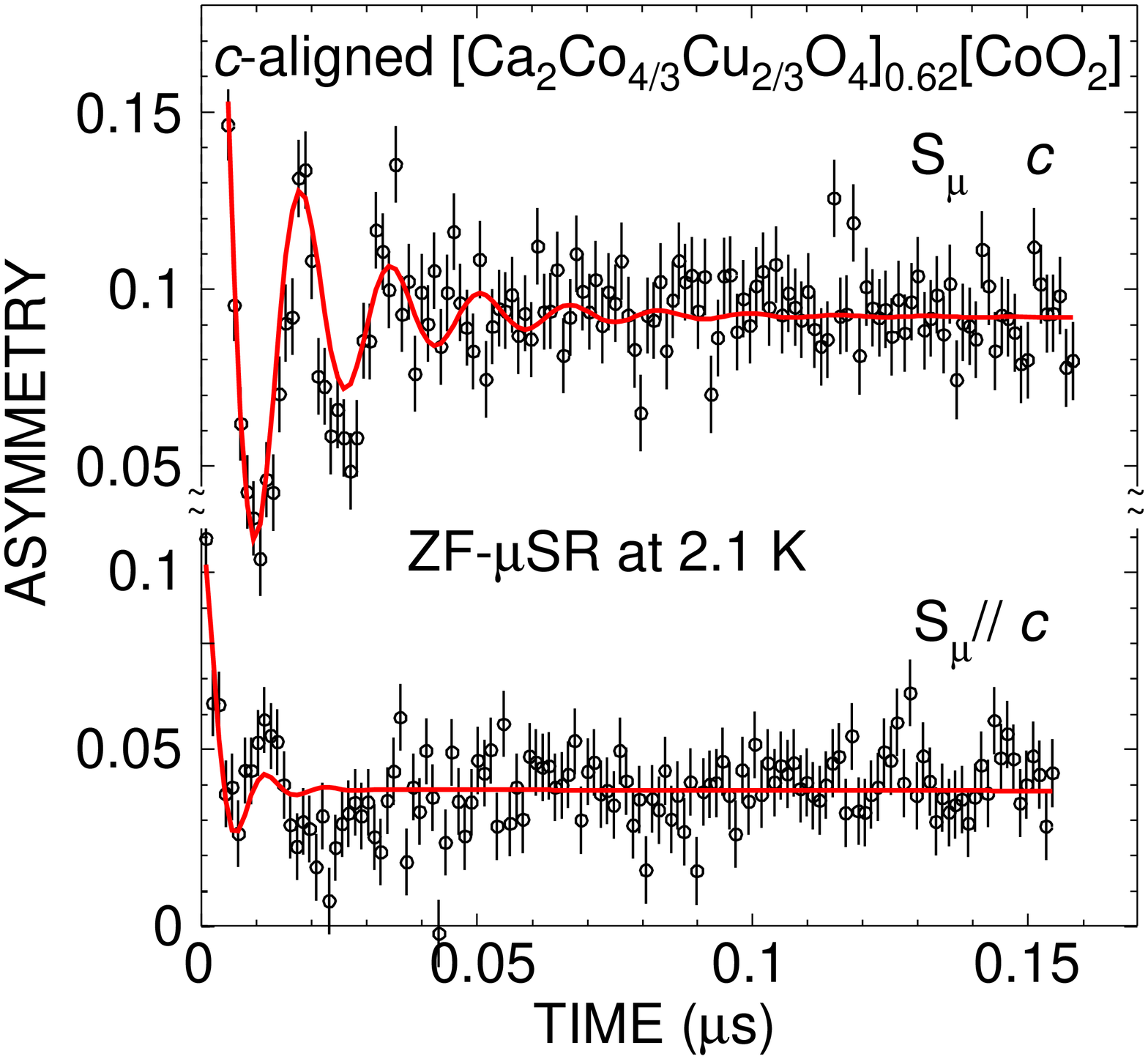}
\caption{\label{fig:ZF-muSR} 
  ZF-$\mu^+$SR time spectra of 
  the $c$-aligned 
  [Ca$_2$Co$_{4/3}$Cu$_{2/3}$O$_4$]$_{0.62}^{\rm RS}$[CoO$_2$] 
  plate at 2.1~K. 
  The configurations of the sample and the muon beam are 
  (top) $\overrightarrow{\bf S}_\mu(0) \perp \mbox{\boldmath $c$}$ and 
  (bottom) $\overrightarrow{\bf S}_\mu(0) \parallel \mbox{\boldmath $c$}$, 
  where $\overrightarrow{\bf S}_\mu(0)$ indicates 
  the initial muon spin direction.  
  }
\end{center}
\end{figure}

Figure~\ref{fig:ZF-muSR} shows ZF-$\mu^+$SR time spectra at 2.1~K 
in the $c$-aligned sample; the top spectrum was obtained with 
the initial $\mu^+$ spin direction $\overrightarrow{\bf S}_\mu(0)$ 
perpendicular to the $c$-axis and the bottom one with 
$\overrightarrow{\bf S}_\mu(0) \parallel \mbox{\boldmath $c$}$.  
A clear oscillation due to quasi-static internal fields 
is observed only when $\overrightarrow{\bf S}_\mu(0) \perp \mbox{\boldmath $c$}$.  
This oscillating spectrum is reasonably well fitted with a combination of 
two zeroth-order Bessel functions of the first kind $J_0$
(for the {\sf IC-SDW})\cite{ICSDW_1,ICSDW_2,ICSDW_3}
and an exponential relaxation function: 
\begin{eqnarray}
 A_0 \, P(t) &=& 
   A_{_{\sf SDW1}} \, J_0(\omega_{\mu 1} t) \, \exp(-\lambda_{\sf SDW1} t)
\cr
 &+& A_{_{\sf SDW2}} \, J_0(\omega_{\mu 2} t) \, \exp(-\lambda_{\sf SDW2} t) 
\cr
 &+& A_{\sf F} \, \exp(-\lambda_{\sf F} t) 
\label{eq:ZFfit}
\end{eqnarray}
\begin{equation}
 \omega_\mu \equiv  2 \pi \nu_\mu = \gamma_{\mu} \; H_{\sf int} 
\label{eq:omg}
\end{equation}
where $A_0$ is the empirical maximum muon decay asymmetry, 
$A_{\sf SDW1}$, $A_{\sf SDW2}$ and $A_{\sf F}$ are the 
asymmetries and 
$\lambda_{\sf SDW1}$, $\lambda_{\sf SDW2}$ and $\lambda_1$ are 
the exponential relaxation rates associated with the three signals.
Also, $\omega_{\mu}$ is the muon precession frequency in the characteristic 
local magnetic field $H_{\sf int}$ due to an {\sf IC-SDW} and
$\gamma_{\mu}$ is muon gyromagnetic ratio.
The two Bessel functions in equation (\ref{eq:ZFfit}) indicate that 
there are two inequivalent muon sites in the
[Ca$_2$Co$_{4/3}$Cu$_{2/3}$O$_4$]$_{0.62}^{\rm RS}$[CoO$_2$] lattice,
because the three signals have a finite value below 140~K. 
Due to some broadening of the {\sf IC-SDW} field distribution, 
the two Bessel functions exhibit an exponential damping. 

We therefore conclude that 
[Ca$_2$Co$_{4/3}$Cu$_{2/3}$O$_4$]$_{0.62}^{\rm RS}$[CoO$_2$] 
undergoes a magnetic transition from a paramagnetic state to 
an {\sf IC-SDW} state, 
similar to the case of [Ca$_2$CoO$_3$]$_{0.62}^{\rm RS}$[CoO$_2$].
\cite{jun1_3L,jun2_3L,jun4_3L}
The absence of a clear oscillation in the bottom spectrum 
of Figure~\ref{fig:ZF-muSR} indicates that 
the internal magnetic field $\overrightarrow{\mbox{\boldmath $H$}}_{\rm int}$ 
is roughly parallel to the $c$ axis \cite{jun4_3L},
since the muon spins 
do not precess in a parallel magnetic field.  
The {\sf IC-SDW} is unlikely to propagate along the $c$ axis 
due both to the two-dimensionality 
and to the misfit between the two subsystems.  
The {\sf IC-SDW} is therefore considered to propagate in the $a$-$b$ plane, 
with oscillating moments directed along the $c$ axis.  

In the [Ca$_2$Co$_{4/3}$Cu$_{2/3}$O$_4$] subsystem, 
1/3 of the Co sites are randomly substituted by Cu \cite{4LCCCO_1}. 
This means that 2 or 3 of the eight of the first nearest neighboring Co ions 
are replaced by Cu for each Co ion in the 
[Ca$_2$Co$_{4/3}$Cu$_{2/3}$O$_4$] subsystem.
Nevertheless, a clear precession was observed 
in the ZF-$\mu^+$SR spectrum below 140~K.  
In addition, the precession frequency ($\sim 60$~MHz) 
at zero temperature is 
almost the same as to 
[Ca$_2$CoO$_3$]$_{0.62}^{\rm RS}$[CoO$_2$] ($\sim 56$~MHz).
Since the long-range order of the Co moments 
in the [Ca$_2$Co$_{4/3}$Cu$_{2/3}$O$_4$] subsystem 
should be strongly hindered by Cu, it is concluded that 
the {\sf IC-SDW} exists not in the 
[Ca$_2$Co$_{4/3}$Cu$_{2/3}$O$_4$] subsystem 
but in the [CoO$_2$] plane.  

\begin{figure}
\begin{center}
\includegraphics[width=8cm]{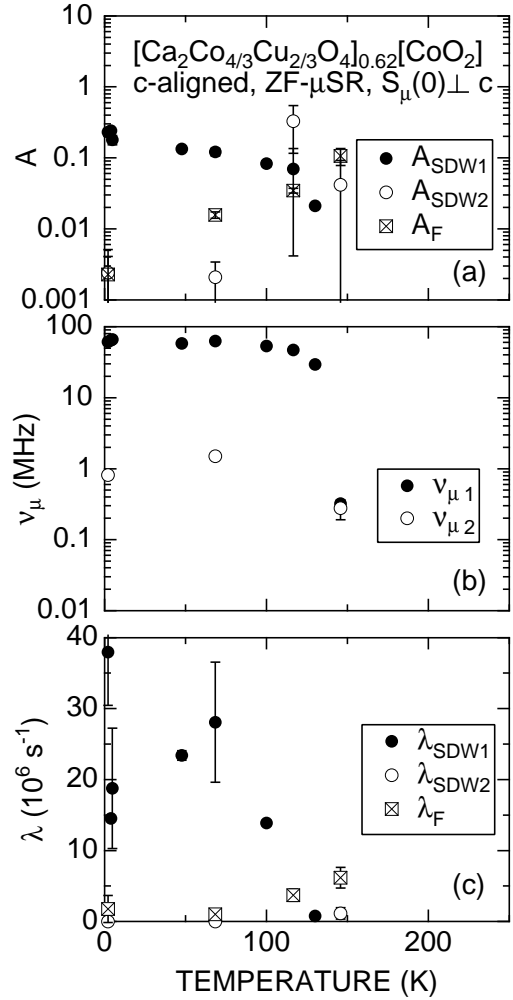}
\caption{\label{fig:SDW1&2} 
 Temperature dependences of 
 (a) $A_{n}$ ($n$ = SDW1, SDW2 and F), 
 (b) $\nu_{\mu n}$ ($n$ = 1 and 2) and 
 (c) $\lambda_{n}$ ($n$ = SDW1, SDW2 and F) 
 for the $c$-aligned 
 [Ca$_2$Co$_{4/3}$Cu$_{2/3}$O$_4$]$_{0.62}^{\rm RS}$[CoO$_2$].
 The data was obtained by fitting of the ZF-$\mu^+$SR time spectra 
 using equation (\ref{eq:ZFfit}).  
  }
\end{center}
\end{figure}

\begin{figure}
\begin{center}
\includegraphics[width=8cm]{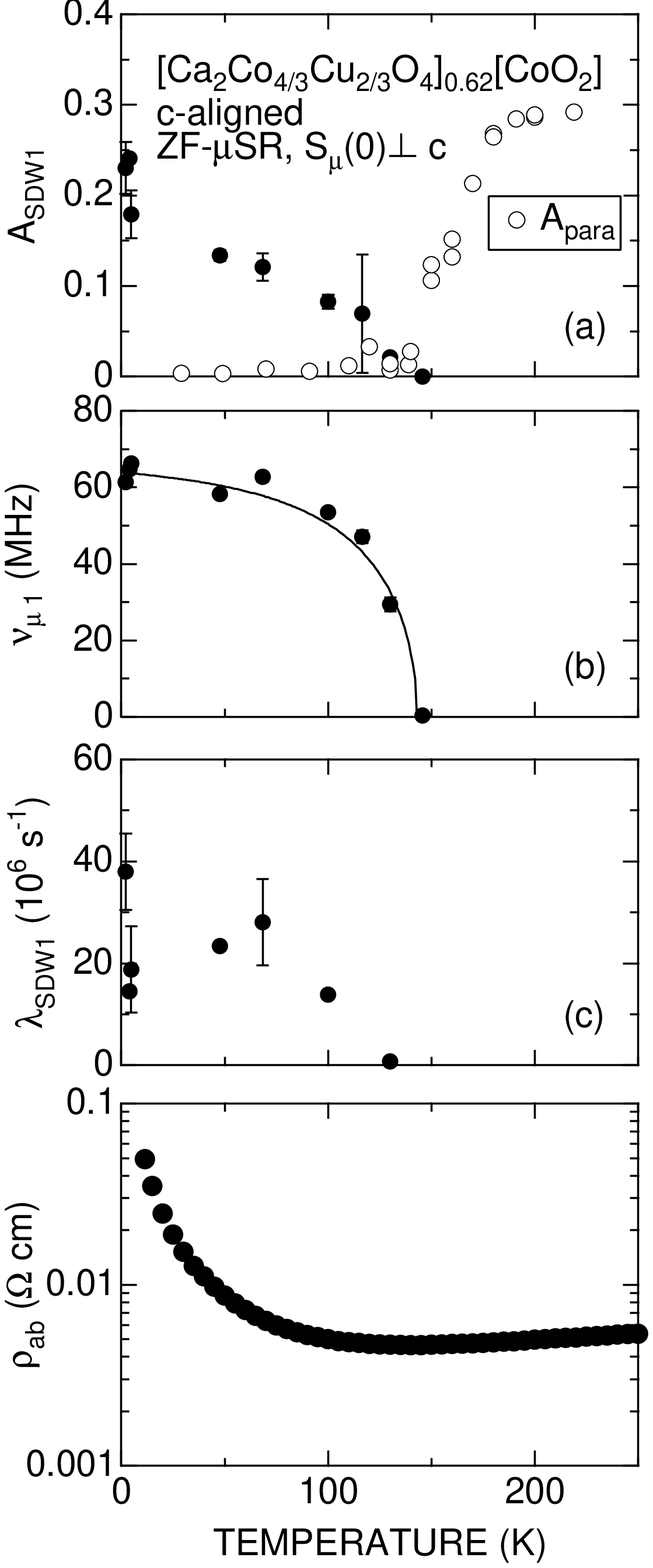}
\caption{\label{fig:SDW1} 
 Temperature dependences of 
 (a) $A_{\sf SDW1}$, 
 (b) $\nu_{\mu 1}$,
 (c) $\lambda_{\sf SDW1}$, and 
 (d) in-plane resistivity ($\rho_{ab}$)  
 for the $c$-aligned 
 [Ca$_2$Co$_{4/3}$Cu$_{2/3}$O$_4$]$_{0.62}^{\rm RS}$[CoO$_2$]. 
 In Figure~\ref{fig:SDW1}(a), $A_{\sf para}$ obtained by 
 the wTF-$\mu^+$SR experiment 
 is also shown for comparison. 
 The solid line in Figure~\ref{fig:SDW1}(b) represents 
 the temperature dependence of the {\sf BCS} gap energy.
 }
\end{center}
\end{figure}

Figures~\ref{fig:SDW1&2}(a) - \ref{fig:SDW1&2}(c) 
show the temperature dependences of 
$A_{n}$ ($n$ = SDW1, SDW2 and F), 
$\nu_{\mu n}$ ($n$ = 1 and 2) and 
$\lambda_{n}$ ($n$ = SDW1, SDW2 and F).
It is clearly seen that the signal associated with the {\sf SDW1} is 
the predominant one among the three signals. 
The volume fraction of the signal from the {\sf SDW1} is 
estimated as $\sim$ 100\% at 2.1~K. 
This suggests that almost all the $\mu^+$ are bound to 
the oxygen ions in the [CoO$_2$] plane, 
with only at the very small portion in the 
[Ca$_2$Co$_{4/3}$Cu$_{2/3}$O$_4$] subsystem.
Therefore, we consider the signal of the {\sf SDW1} and 
ignore the contribution from the other signals.

Figures~\ref{fig:SDW1}(a) - \ref{fig:SDW1}(d) show 
the temperature dependences of 
$A_{\sf SDW1}$, $\nu_{\mu 1}$, $\lambda_{\sf SDW1}$, and 
in-plane resistivity ($\rho_{ab}$). 
In particular, $A_{\sf SDW1}$ increases monotonically 
with decreasing $T$ from 140~K, 
although $A_{\sf para}$ obtained by the wTF-$\mu^+$SR measurement 
exhibits a rapid decrease below 200~K and levels off to almost 0 below 140~K 
(see Figure~\ref{fig:SDW1}(a)). 
This suggests that a long-range {\sf IC-SDW} order completes 
below $\sim$ 140~K (= $T_{\sf SDW}$), 
while a short-range order appears below 200~K (= $T_{\sf SDW}^{\rm on}$), 
as in the case of 
[Ca$_2$CoO$_3$]$_{0.62}^{\rm RS}$[CoO$_2$] \cite{jun3_3L,jun4_3L}.  
Actually, this is in good agreement with 
the temperature dependence of $\rho_{ab}$; 
that is, $\rho_{ab} (T)$ is metallic above 140~K 
and semiconducting below 140~K.
On the other hand, there was no clear anomalies around 140~K 
in the $\chi(T)$ curve (see Fig.~\ref{fig:chi}) 
probably due to effects of the grain boundaries,
a possible inhomogeneous distribution of the cations and oxygen ion,    
and magnetic anisotropy. 
Indeed, the $\chi(T)$ curve for the
[Ca$_2$CoO$_3$]$_{0.62}^{\rm RS}$[CoO$_2$] single crystals 
exhibited a clear but small maximum at 27~K indicating 
the formation of the {\sf IC-SDW} order 
only when $H \parallel c$\cite{jun4_3L}, 
whereas those for the $c$-aligned and random polycrystalline samples did not. 
Therefore, if a large single crystal of 
[Ca$_2$Co$_{4/3}$Cu$_{2/3}$O$_4$]$_{0.62}^{\rm RS}$[CoO$_2$]
is available, 
its $\chi_{\rm c}(T)$ curve would also have an anomaly around 140~K.

The $\nu_{\mu 1}(T)$ curve is well explained by 
the energy gap function in the {\sf BCS} weak-coupling theory 
(see Figure~\ref{fig:SDW1}(b)), 
as expected for the {\sf IC-SDW} state \cite{ICSDW_4}. 
As seen in Figure~\ref{fig:SDW1}(c), $\lambda_{\sf SDW1}$ seems to 
increase monotonically with decreasing $T$. 
The exponential damping of the {\sf IC-SDW} oscillation 
is most likely to be caused by the misfit between the 
[Ca$_2$Co$_{4/3}$Cu$_{2/3}$O$_4$] and the [CoO$_2$] 
subsystem along the $b$ axis. 
Since we used the $c$-aligned sample for 
the current ZF-$\mu^+$SR measurements, 
the result obtained with $\overrightarrow{\bf S}_\mu(0) \perp \mbox{\boldmath $c$}$ is 
the average information along the $a$ and $b$ axis. 
Thus, the anisotropic {\sf IC} modulation in the $a-b$ plane 
is considered to be the origin of the broadening 
of the {\sf IC-SDW} field distribution 
at the $\mu^+$ sites.

\subsection{Pure and Sr, Y and Bi doped 
[Ca$_2$CoO$_3$]$_{0.62}^{\rm RS}$[CoO$_2$]}

\begin{figure}
\begin{center}
\includegraphics[width=8cm]{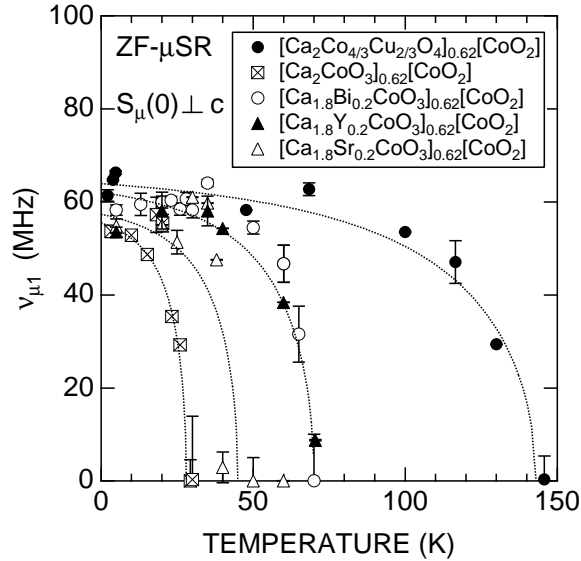}
\caption{\label{fig:SDW_others} 
 Temperature dependences of $\nu _{\mu 1}$ for the $c$-aligned 
 [Ca$_2$Co$_{4/3}$Cu$_{2/3}$O$_4$]$_{0.62}^{\rm RS}$[CoO$_2$]
 and $c$-aligned pure and doped  
 [Ca$_2$CoO$_3$]$_{0.62}^{\rm RS}$[CoO$_2$].
 The dotted lines represent 
 the temperature dependence of the {\sf BCS} gap energy. 
  }
\end{center}
\end{figure}

Similar anisotropic ZF-$\mu^+$SR spectra were also observed for 
the $c$-aligned 
[Ca$_2$CoO$_3$]$_{0.62}^{\rm RS}$[CoO$_2$] 
below $\sim$ 30~K \cite{jun4_3L}, 
and the ZF-$\mu^+$SR time spectra 
with $\overrightarrow{\bf S}_\mu(0) \perp \mbox{\boldmath $c$}$ 
were well fitted using the Bessel function, 
$A_{\sf SDW1}J_0(\omega_{\mu 1} t)$exp($\lambda _{\sf SDW1}$).

Figure~\ref{fig:SDW_others} shows the $\nu _{\mu 1}(T)$ curve 
for the $c$-aligned pure and doped 
[Ca$_2$CoO$_3$]$_{0.62}^{\rm RS}$[CoO$_2$]
and 
[Ca$_2$Co$_{4/3}$Cu$_{2/3}$O$_4$]$_{0.62}^{\rm RS}$[CoO$_2$]
samples.
The data were obtained by fitting the ZF-$\mu^+$SR time spectra 
with $\overrightarrow{\bf S}_\mu(0) \perp \mbox{\boldmath $c$}$ 
using equation~(\ref{eq:ZFfit}). 
All the temperature dependences of $\nu_{\mu 1}$ 
for the $c$-aligned layered cobaltites are 
well fitted by the {\sf BCS} weak-coupling theory. 
It should be noted that all the samples show 
approximately the same precession frequency at zero temperature, 
although the transition temperatures are different 
from 27 to $\sim$140~K; {\sl i.e.},
$T_{\sf SDW}$ = 27~K for 
[Ca$_3$CoO$_3$]$_{0.62}^{\rm RS}$[CoO$_2$],
$\sim$ 45~K for  
[Ca$_{1.8}$Sr$_{0.2}$CoO$_3$]$_x^{\rm RS}$[CoO$_2$],
$\sim$ 80~K for  
[Ca$_{1.8}$Y$_{0.2}$CoO$_3$]$_x^{\rm RS}$[CoO$_2$] 
and
[Ca$_{1.8}$Bi$_{0.2}$CoO$_3$]$_x^{\rm RS}$[CoO$_2$] 
and $\sim$ 140~K for 
[Ca$_2$Co$_{4/3}$Cu$_{2/3}$O$_4$]$_{0.62}^{\rm RS}$[CoO$_2$].
This suggests that the local magnetic field $H_{\rm int}$(0~K) 
is independent of both dopant and the number of layers 
in the rocksalt-type subsystem, 
if we assume that the muon sites are essentially the same 
in all these compounds.  
This supports the conclusion that 
the {\sf IC-SDW} exists not in the rocksalt-type subsystem 
but in the [CoO$_2$] plane and, as a result,
the $\mu^+$ sites are considered to be the vicinity of 
the O ions in the [CoO$_2$] plane. 

\section{Discussion}

\subsection{{\sf IC-SDW}: common to all the (good thermoelectric) cobaltites ?}
Although the possible $\mu^+$ sites are bound to the oxygen ions in the [CoO$_2$] plane and 
the two inequivalent oxygen ions in the [Ca$_2$Co$_{4/3}$Cu$_{2/3}$O$_4$] subsystem,
the $\mu^+$ sites are most likely to the vicinity of the O ions in the [CoO$_2$] plane. 
We, therefore, calculate the dipole field at the O ions in the [CoO$_2$] plane at first.
The bond length $d$ of Co-O in the [CoO$_2$] plane is 0.184~nm 
in [Ca$_2$Co$_{4/3}$Cu$_{2/3}$O$_4$]$_{0.62}^{\rm RS}$[CoO$_2$] \cite{4LCCCO_2},
whereas it is 0.197~nm in 
[Ca$_2$CoO$_3$]$_{0.62}^{\rm RS}$[CoO$_2$] \cite{CCO_4}. 
A simple estimate of dipole field from one ion is given by;
\begin{eqnarray}
 H_{\rm int} &=& \frac{m_{\rm d}}{4\pi\mu _0d^3} ,
\label{eq:dipole}
\end{eqnarray}
where $m_{\rm d}$ is 
the dipole moment of the Co ions in the {\sf IC-SDW} state and 
$\mu _0$ is the permeability of free space. 
Using the above $d$ values and the observed $H_{\rm int}$, 
we estimate $m_{\rm d} = 3.2~\mu _{\rm B}$ for 
[Ca$_2$Co$_{4/3}$Cu$_{2/3}$O$_4$]$_{0.62}^{\rm RS}$[CoO$_2$] 
and 3.3~$\mu _{\rm B}$ for 
[Ca$_2$CoO$_3$]$_{0.62}^{\rm RS}$[CoO$_2$] (see Table~1).
\begin{table}
\caption{\label{tab:table1}Internal magnetic field determined by $\mu^+$SR and 
the corresponding magnetic moment 
at the Co site in the [CoO$_2$] plane estimated by equation~(\ref{eq:dipole}). 
The values of $d_{\rm Co-O}$ in the rocksalt-type 
subsystem are 0.213-0.230~nm in 
[Ca$_2$Co$_{4/3}$Cu$_{2/3}$O$_4$]\cite{4LCCCO_2} 
and 0.179-0.228~nm in 
[Ca$_2$CoO$_3$]\cite{CCO_4}.
}
\begin{indented}
\item[]\begin{tabular}{@{}ccc}
\br
cobaltite & [Ca$_2$Co$_{4/3}$Cu$_{2/3}$O$_4$]$_{0.62}^{\rm RS}$[CoO$_2$] & [Ca$_2$CoO$_3$]$_{0.62}^{\rm RS}$[CoO$_2$] \\
\mr
$\nu_{\mu,1}(0~K)$~(MHz) & 63.9 & 55.5 \\
$H_{\rm int}$~(kOe) & 4.71 & 4.08 \\
$d_{\rm Co-O}$ in [CoO$_2$]~(nm) & 0.184 & 0.197 \\
$m_{\rm d}~(\mu_{\rm B})$ & 3.2 & 3.3\\
\br
\end{tabular}
\end{indented}
\end{table}
If we ignore the effect of the distortion of the CoO$_6$ octahedra 
in the [CoO$_2$] plane, 
the number of the nearest neighboring Co ions for the O ion is three
\cite{4LCCCO_1}; 
hence, the ordered Co moment in the {\sf IC-SDW} state is therefore 
roughly estimated as $\sim 1.1~\mu _{\rm B}$/Co ion for both compounds. 
This is in good agreement with the amplitude of the {\sf IC-SDW} 
estimated by the mean field theory 
($\sim 0.86~\mu _{\rm B}$/Co ion) \cite{jun1_3L}.
Here, it should be noted that the muon locates
probably $\sim$0.1~nm away from the oxygen ions, 
and that there is no space for it 
in the CoO$_6$ octahedra in the [CoO$_2$] plane 
as in the case for the high-$T_{\rm c}$ cuprates\cite{ICSDW_3}. 
Thus, the accuracy of the above estimation is very limited; 
{\sl i.e.}, $m_{\rm d}$=0.3-11.7~$\mu _{\rm B}$ for
[Ca$_2$Co$_{4/3}$Cu$_{2/3}$O$_4$]$_{0.62}^{\rm RS}$[CoO$_2$], 
if $d_{\rm Co-\mu}=0.184\pm0.1$~nm. 
In order to determine the $\mu^+$ sites, 
both further experiments on the layered cobaltites and 
a theoretical research are necessary. 

There are two Co sites in the 
[Ca$_2$Co$_{4/3}$Cu$_{2/3}$O$_4$]$_{0.62}^{\rm RS}$[CoO$_2$] 
lattice. 
Thus, it is difficult to determine the Co valence in the [CoO$_2$] 
plane by a $\chi$ measurement or a chemical titration technique alone.
Here, we assume that the magnitude of $S$ depends on 
the concentration $y$ of Co$^{4+}$ ions in the [CoO$_2$] plane as
\cite{koshibae_1};
\begin{eqnarray}
 S_{T\rightarrow\infty} &=&
 -\frac{k_{\rm B}}{e} {\rm ln}\biggl(\frac{g_3}{g_4}\frac{y}{1-y}\biggr) ,
\label{eq:tep}
\end{eqnarray}
where $k_{\rm B}$ is the Boltzmann constant, $e$ is the elementary charge 
and $g_3$ and $g_4$ are the multiplying numbers of the spin configurations 
of Co$^{3+}$ and Co$^{4+}$, respectively. 
Since the electron configurations of both Co$^{3+}$ and Co$^{4+}$ 
are in the low-spin state ($t_{2g}^6$ and $t_{2g}^5$) 
\cite{CCO_2,CCO_3,jun1_3L,TlSCO_1,4LBiSrCO_2},  
$g_3$=1 and $g_4$=6. 
Then, using the value of $S$(300~K) $\sim 150~\mu$VK$^{-1}$, 
we obtain $y \sim$ 0.51, {\sl i.e.} 
the average valence of the Co ions in the [CoO$_2$] plane is +3.51. 
This indicates that almost the same amounts of Co$^{3+}$ and Co$^{4+}$ 
coexist in the [CoO$_2$] plane. 
In other words, the Co spin ({\bf S}=1/2) occupies about half of the sites of
the two-dimensional-triangular lattice of Co ions. 
This is significant to achieve the {\sf IC-SDW} long-range order 
in the triangular lattice. 
In addition, the average Co moment is calculated as 
$\sim 0.86~\mu _{\rm B}$/Co ion, 
which is consistent with the value estimated above.

It is worth noting that $\mu^+$ sites are bound to the O ions
in the [CoO$_2$] plane. 
This means that the $\mu^+$ feel mainly the magnetic field 
in the [CoO$_2$] plane. 
Thus, the {\sf IC-SDW} is most unlikely to be caused 
by the misfit between the two subsystems, 
but to be an intrinsic behavior of the [CoO$_2$] plane. 
Indeed, the recent $\mu^+$SR experiments on 
[Na]$_x$[CoO$_2$], 
which consist of the alternating stack of Na and [CoO$_2$] planes, 
also indicate the existence of 
a commensurate {\sf SDW} or a ferrimagnetic state 
below 22~K for [Na]$_{0.75}$[CoO$_2$] \cite{jun5_1L}
and an {\sf IC-SDW} state below 19~K 
for [Na]$_{0.9}$[CoO$_2$] \cite{jun6_1L}.
Therefore, it is concluded that the {\sf IC-SDW} state is 
a common behavior for the layered cobaltites, 
although the magnitude of $T_{\rm SDW}$ depends on 
the Co valence in the [CoO$_2$] plane and 
the structure of the subsystem sandwiched by the two [CoO$_2$] planes 
(see Figure~\ref{fig:SDW_others}).

The {\sf IC-SDW} order in 
the two layered cobaltites
is assigned to be a spin ({\bf S}=1/2) order 
on the two-dimensional triangular lattice 
({\sl i.e.} the CoO$_2$ plane) 
at non-half filling.
Such a problem was investigated by several workers 
using the Hubbard model within a mean field approximation
\cite{MHonTL_1,MHonTL_2,MHonTL_3};
\begin{eqnarray}
 {\cal H}&=&-t\sum_{<ij>\sigma}c_{i\sigma}^{\dagger}c_{j\sigma} + 
 U\sum_i n_{i\uparrow}n_{i\downarrow} ,
\label{eq:Hubbard}
\end{eqnarray}
where $c_{i\sigma}^{\dagger}(c_{j\sigma})$ creates (destroys) 
an electron with spin $\sigma$ on site $i$, 
$n_{i\sigma}=c_{i\sigma}^{\dagger}c_{i\sigma}$ is the number operator, 
$t$ is the nearest-neighbor hopping amplitude and 
$U$ is the Hubbard on-site repulsion.
The electron filling $n$ is defined as $n$ = (1/2$N$)$\sum_i^N n_i$,
where $N$ is the total number of sites.  

At $T$=0 and $n$=0.5 ({\sl i.e.}, 
the average valence of the Co ions in the [CoO$_2$] plane is +4),
as $U$ increased from 0, the system is 
paramagnetic metal up to $U/t \sim 3.97$, 
and changes into a metal 
with a spiral {\sf IC-SDW}, 
and then at $U/t \sim 5.27$, 
a first-order metal-insulator transition occurs \cite{MHonTL_1}.
Also the calculations predict that
\cite{MHonTL_2, MHonTL_3}, 
as $n$ increases from 0, 
the magnitude of $U/t$ at the boundary 
between the paramagnetic and {\sf SDW} phases 
decreases with increasing slope 
($d(U/t)/dn$) up to $n$=0.75. 
Even for $U/t$=0, the {\sf SDW} phase is stable 
at $n$=0.75.
$U/t$ increases with further increasing $n$,
with decreasing slope. 
In other words, at $n$=0.75 ({\sl i.e.} 
the average valence of the Co ions in the [CoO$_2$] plane is +3.5),
a spiral {\sf IC-SDW} state is expected to appear at the highest temperature 
in the $n$ range between 0.5 and 1 \cite{MHonTL_2}. 

The value of $n$ is estimated as 0.74 for 
[Ca$_2$Co$_{4/3}$Cu$_{2/3}$O$_4$]$_{0.62}^{\rm RS}$[CoO$_2$] 
and 0.715, 0.70, 0.73 and 0.73 for pure, Sr-, Y-, and Bi-doped 
[Ca$_2$CoO$_3$]$_{0.62}^{\rm RS}$[CoO$_2$] 
respectively, using eq.~(\ref{eq:tep}) and $S$(300~K).
Therefore, the fact that $T_{\sf SDW}$ for 
[Ca$_2$Co$_{4/3}$Cu$_{2/3}$O$_4$]$_{0.62}^{\rm RS}$[CoO$_2$] 
is higher than those for pure and doped 
[Ca$_2$CoO$_3$]$_{0.62}^{\rm RS}$[CoO$_2$] 
(see Figure~\ref{fig:SDW_others}) 
is roughly explained by the model calculations, 
if we ignore the data for Sr-doped 
[Ca$_2$CoO$_3$]$_{0.62}^{\rm RS}$[CoO$_2$]. 

Also, it should be pointed out that 
the increased two-dimensionality induced by the increase in 
the interlayer distance between CoO$_2$ planes 
plays a significant role to increase $T_{\sf SDW}$. 
This is because $T_{\sf SDW}$ for 
[Ca$_2$Co$_{4/3}$Cu$_{2/3}$O$_4$]$_{0.62}^{\rm RS}$[CoO$_2$] 
is considerably higher than those for pure and doped 
[Ca$_2$CoO$_3$]$_{0.62}^{\rm RS}$[CoO$_2$].
Here, we repeat the prediction of the model calculations. 
That is, the {\sf SDW} phase is stable at $n$=0.75 even for $U/t$=0; 
this means that, on the ideal two-dimensional triangular lattice at $n$=0.75, 
the {\sf SDW} spin structure appears at temperatures below its melting point. 
Indeed, the large observed transition width (60~K) are consistent with 
enhanced two-dimensionality and resulting spin fluctuations. 
Therefore, not only the shift of $n$ towards the optimal value (0.75) 
but also the enhanced two-dimensionality 
are considered to increase $T_{\sf SDW}$ for 
[Ca$_2$Co$_{4/3}$Cu$_{2/3}$O$_4$]$_{0.62}^{\rm RS}$[CoO$_2$]. 

Since the magnitude of $\nu_{\mu 1}$ indicates 
the amplitude of the {\sf IC-SDW} directly, 
there are two predominant factors for $\nu_{\mu 1}$. 
That is, the magnitude and direction of the Co spin in the [CoO$_2$] plane.
The former is a constant at low temperatures, 
because the spin configurations for 
Co$^{3+}$ and Co$^{4+}$ ions 
({\bf S}=0 and 1/2) are not affected by 
both dopant and the number of layers 
in the rocksalt-type subsystem.  
The latter is a function of magnetic anisotropy in the lattice. 
Here the magnetic anisotropy in
[Ca$_2$Co$_{4/3}$Cu$_{2/3}$O$_4$]$_{0.62}^{\rm RS}$[CoO$_2$] 
is most likely to be equivalent to that in 
[Ca$_2$CoO$_3$]$_{0.62}^{\rm RS}$[CoO$_2$]
due to similar environment 
around the [CoO$_2$] plane in the two cobaltites.
That is, the first and second nearest adjacent plane
for the [CoO$_2$] plane are
the Ca-O and (Co$_{2/3}$Cu$_{1/3}$)-O or Co-O plane 
in the rocksalt-type subsystem,
although the third nearest plane is different for the two cobaltites.
Therefore, $\nu_{\mu 1}$(0~K) is considered to be approximately same 
for all the cobaltites investigated here (see Fig.~\ref{fig:SDW_others}).
 
\subsection{The anomaly at $\sim$ 85~K in the susceptibility}

The $\chi(T)$ curve exhibits a clear anomaly at $\sim$ 85~K, 
while the results of both wTF-$\mu^+$SR and ZF-$\mu^+$SR do not 
(see Figures~\ref{fig:wTF-muSR}, \ref{fig:SDW1&2} and \ref{fig:SDW1}).
On the other hand, there is no marked anomaly 
at 140 - 200~K in the $\chi(T)$ curve 
(see Figure~\ref{fig:chi}),
although $\mu^+$SR detects the formation of the {\sf IC-SDW} state.
This is quite similar to the case for 
[Ca$_2$CoO$_3$]$_{0.62}^{\rm RS}$[CoO$_2$];
where according to the $\mu^+$SR experiments, 
it was found that
the short-range {\sf IC-SDW} order appeared 
below $\sim$ 100~K and 
the long-range order completed below 27~K, 
whereas the $\chi(T)$ curve below 300~K only exhibited 
a ferrimagnetic transition at 19~K
\cite{jun1_3L,jun2_3L,jun3_3L,jun4_3L}.

The structures of the {\sf IC-SDW} of both compounds 
are considered to be essentially the same, as discussed above.
For [Ca$_2$CoO$_3$]$_{0.62}^{\rm RS}$[CoO$_2$],
the {\sf IC-SDW} is induced 
by ordering of the Co moments in the [CoO$_2$]
plane, whereas the 
ferrimagnetic ordering arises because of 
the interlayer coupling 
between the Co spins in the [CoO$_2$] and
[Ca$_2$CoO$_3$] subsystems\cite{jun3_3L,jun4_3L}.

As seen in Figure~\ref{fig:SDW1}(b), 
the microscopic internal {\sf IC-SDW} field 
does not change around 85~K.
Furthermore, the extremely small value of the asymmetry of 
the exponential relaxation function $A_{\sf F}$, even at 2.1~K, 
(see Figure~\ref{fig:SDW1}(a)) suggests that 
the anomaly at $\sim$ 85~K is not a transition to a spin-glass state
\cite{ICSDW_2,ICSDW_3,SG_1}. 
Therefore, the anomaly at $\sim$ 85~K  
is most likely caused by some order between the two subsystems, 
{\sl i.e.},
the [Ca$_2$Co$_{4/3}$Cu$_{2/3}$O$_4$] and the [CoO$_2$]. 
The interlayer coupling between both subsystems 
is expected to be basically antiferromagnetic ({\sf AF}), 
because there was no clear $M-H$ loop even at 5~K.
In other words, a two-dimensional {\sf AF} ({\sf IC-SDW}) order 
of the Co spins completes at 140~K, 
whereas a three-dimensional {\sf AF} order occurs below $\sim$ 85~K. 

The [Ca$_2$Co$_{4/3}$Cu$_{2/3}$O$_4$] subsystem consists of 
two Ca-O planes and two (Co$_{2/3}$Cu$_{1/3}$)-O planes; 
so that, the two (Co$_{2/3}$Cu$_{1/3}$)-O planes are sandwiched by 
the two Ca-O planes. 
Considering the fact that 1/3 of the Co ions are replaced by Cu ions 
and that the thickness of the rocksalt-type subsystem increases 
due to the extra (Co$_{2/3}$Cu$_{1/3}$)-O plane, 
the coupling along the $c$-axis for 
[Ca$_2$Co$_{4/3}$Cu$_{2/3}$O$_4$]$_{0.62}^{\rm RS}$[CoO$_2$] 
is rather weak compared with that for 
[Ca$_2$CoO$_3$]$_{0.62}^{\rm RS}$[CoO$_2$]. 
This weak coupling is likely to be a significant factor 
for the thermal hysteresis of the $\chi(T)$ curve. 
That is, the {\sf IC-SDW} in each [CoO$_2$] plane 
is considered to be moved by the external magnetic field 
($\sim$ 10~kOe) at 85 - 140~K, but fixed below $\sim$ 85~K.

\section{Summary}
In order to elucidate the magnetism in 'good' thermoelectric layered cobaltites, 
$\mu^+$SR spectroscopy has been used 
on a $c$-aligned polycrystalline 
[Ca$_2$Co$_{4/3}$Cu$_{2/3}$O$_4$]$_{0.62}^{\rm RS}$[CoO$_2$] 
sample at temperatures below 300~K.  
It was found that 
[Ca$_2$Co$_{4/3}$Cu$_{2/3}$O$_4$]$_{0.62}^{\rm RS}$[CoO$_2$] 
exhibits a transition at around 140~K from a paramagnetic 
to an incommensurate spin density wave {\sf IC-SDW} state, 
although a short-range order appears below $\sim$ 180~K. 

By comparison with the $\mu^+$SR results on pure and doped 
[Ca$_2$CoO$_3$]$_{0.62}^{\rm RS}$[CoO$_2$], 
the {\sf IC-SDW} appears to be common behavior for these cobaltites.
The characteristic features of the {\sf IC-SDW} are as follows;
\begin{enumerate}
\item 
A long-range {\sf IC-SDW} order completes below $T_{\sf SDW}$, 
while a short-range order is observed at 40 - 60~K higher than $T_{\sf SDW}$ 
with a transition width $\Delta T$ = 40 - 60~K.
\item 
The {\sf IC-SDW} propagates in the $a$-$b$ plane, 
with oscillating moments directed along the $c$ axis.  
\item 
The {\sf IC-SDW} exists not in the rocksalt-type subsystem but in the [CoO$_2$] plane.
\end{enumerate}
The magnitude of $T_{\rm SDW}$ is found to be sensitive both 
to the Co valence in the [CoO$_2$] plane, 
{\sl i.e.}, the occupancy of the Co spin ($S$=1/2) in the triangular lattice, 
and to the structure of the subsystem sandwiched by the two [CoO$_2$] planes. 
Therefore, physical properties of the layered cobaltites should be investigated systematically as functions of the Co valence in the 
[CoO$_2$] plane and the distance between the two adjacent [CoO$_2$] planes.

\ack
We thank Dr. S.R. Kreitzman and Dr. D.J. Arseneau of TRIUMF
for help with the $\mu^+$SR experiments. 
Also, we thank Mr. A. Izadi-Najafabadi and Mr. S.D. LaRoy 
of University of British Columbia for help with the experiments. 
We appreciate useful discussions with 
Dr. R. Asahi of Toyota CRDL, 
Professor U. Mizutani, Professor H. Ikuta and 
Professor T. Takeuchi of Nagoya University.  
This work was supported 
at Toyota CRDL by joint research and development with
International Center for Environmental Technology Transfer in 2002-2004,
commissioned by the Ministry of Economy Trade and Industry of Japan,
at UBC by the Canadian Institute for Advanced Research, 
the Natural Sciences and Engineering Research Council of Canada, 
and at TRIUMF by the National Research Council of Canada.


\end{document}